\documentclass[universe,communication,accept,moreauthors,pdftex,trackchanges]{Definitions/mdpi}

\firstpage{1}
\pubvolume{8}
\issuenum{3}
\articlenumber{159}
\pubyear{2022}
\copyrightyear{2022}
\externaleditor{Academic Editor:Andrea Melandri	and Silvia Piranomonte}

\datereceived{11 February 2022}
\dateaccepted{1 March 2022}
\datepublished{2 March 2022}
\hreflink{https://doi.org/10.3390/universe8030159} 

\Title{A Comprehensive Study of Bright Fermi-GBM Short Gamma-Ray Bursts: I. Multi-Pulse Lightcurves and Multi-Component Spectra}

\TitleCitation{A Comprehensive Study of Bright Fermi-GBM Short Gamma-Ray Bursts: I. Multi-Pulse Lightcurves and Multi-Component Spectra}

\Author{Peng-Wei Zhao 
 and Qing-Wen Tang *\orcidA{}}

\AuthorNames{Peng-Wei Zhao and Qing-Wen Tang}

\AuthorCitation{Zhao, P.-W.; Tang, Q.-W.}

\address [1]{%
Department of Physics, Nanchang University, Nanchang 330031, China; zhaopengwei5@email.ncu.edu.cn}

\corres{\hangafter=1 \hangindent=1.05em \hspace{-0.82em}Correspondence: qwtang@ncu.edu.cn }

\abstract{Sorted by the photon fluences of short Gamma-ray Bursts (SGRBs) detected by the Fermi- Gamma Ray Burst Monitor (GBM), nine brightest bursts are selected to perform a comprehensive analysis. All GRB lightcurves  {are fitted} well by 1 to 3 pulses that are modelled by fast-rising exponential decay profile (FRED), within which the resultant rising time is strongly positive-correlated with the full time width at half maxima (FWHM). A photon spectral  {model involving} a cutoff power-law function and a standard blackbody function (CPL + BB) could reproduce the spectral energy distributions of these SGRBs well in the bursting phase. The CPL's peak energy is found strongly positive-correlated with the BB's temperature, which indicates they might be from the same physical origin. Possible physical origins are discussed to account for these correlations.}


\begin{document}

\section{Introduction}

{Gamma-ray Bursts (GRBs) are the most energetic transient events in the universe, which is only after the bing bang.} Short Gamma-ray Burst (SGRB)  {is an important astrophysical phenomena} to shed light on the properties of the stellar physics, especially on the merger events of two compact objects in the gravitational-wave era, such as two neutron stars (NS-NS), which is firstly proved in SGRB 170817A~\citep{Abbott2017}. During the short emission timescale of such event, i.e., less than two seconds in the prompt phase, SGRB typically consists of several pulses in its lightcurve (LC) and multiple spectral components in its spectral energy distribution (SED). The radiation process accounted  {for these properties} during the prompt emission of SGRBs is still ambiguous, which however  {can revel} some aspects of GRB, such as the central engine, the outflow composition and the structures of the relativistic jet~ ~\citep{Duffell2018,Mooley2018a,Mooley2018b, Alexander2018,Granot2018}. For pulses of those LCs, it is found that the rise time width and the full time width are positively correlated with each other  {in most of the long GRBs (LGRBs) detected by the Burst} and Transient Source Experiment (BATSE)  {on board Compton} Gamma Ray Observatory (CGRO)~\citep{Kocevski2003,Lu2006,Peng2012}. However, such pulses in most BATSE SGRBs have not enough high count rates to be decomposed. For the SEDs represented by the nonthermal models in most SGRBs~\citep{Band1993}, two bright SGRBs (GRB 120323A and GRB 170206A) with the standard blackbody component detection both in their time-integrated and time-resolved spectra are reported in the literatrue~\citep{Guiriec2013,Zhao2022}, while the multiple blackbody component (mBB, a non-standard thermal component) is found in several redshift-measured SGRBs~\citep{Iyyani2021}. In order to discuss the possible physical origins in SGRBs, 9 fluence-selected brightest SGRBs amongst 522 Fermi-Gamma Burst Monitor (GBM) as of 2021 December are selected to perform the comprehensive analysis, in which both the LCs and SEDs are fitted. In Section~\ref{method}, we describe the method. The results and possible correlaions are presented in Section~\ref{result}. The discussion on these correlations is presented in Section~\ref{discussion}. We present the summary and conclusion in Section~\ref{conclusion}.

\section{Data Analysis \label{method}}
{In this section, we present the sample selection and the general method for the lightcurve fitting and spectral energy distribution fitting, employing the Fermi/GBM observations.}
\subsection{Sample Selection}
Fermi/GBM has two types of  {scintillation} detectors, such as 12  {Sodium Iodide (NaI) units} named from `n0' to `n9', `na' and `nb', and 2  {Bismuth Germanate (BGO) units} named `b0' and `b1'. NaIs cover the photon energy between about 8 keV and 1 MeV while BGOs between about 200 keV and 40 MeV.
{Among 522 SGRBs detected by the Fermi-GBM} as of 2021 December,  {nine are} selected with 50--300 keV energy fluence $S_{\rm 50-300\ keV}$ above \mbox{7$\times 10^{-6}$ ${\rm erg\ cm^{-2}}$} and GBM $T_{90}$ less than 2 s. GBM $T_{90}$ is between GBM $T_{05}$ and $T_{95}$, that are the times when 5$\%$ and 95$\%$ of the total GRB energy fluence is accumulated respectively. For each GRB, we selected four detectors in the following analysis, which are  {closest} to the best-localizaion GRB position as shown in the Table ~\ref{tab:info}. Their data can be downloaded from the public data site of Fermi/GBM.

\begin{table}[H]
	\caption{GRB sample.\label{tab:info}}
	\begin{tabular}{m{2.5cm}<{\centering}cccm{3cm}<{\centering}m{3.2cm}<{\centering}}
		\toprule
		\textbf{GRB}& \boldmath{$T_{90}$}  &\boldmath{ $T_{05}$ } & \boldmath{$T_{95}$}  	&\boldmath{ \textbf{Detetor}} &\boldmath{$S_{\rm 50-300\ keV}$}\\
	   &   \textbf{(s)}&   \textbf{(s)}&   \textbf{(s)}	&   & \textbf{(}\boldmath{$10^{-6}$ ${\rm erg\ cm^{-2}}$}\textbf{)}\\
		\midrule
		090227B		& 0.304		& $-$0.016    & 0.288		& `n0','n1','n2','b0' &11.1$~\pm~$0.1\\
		120323A		& 0.384		& 0       & 0.384		& `n0',`n3',`n4',`b0'&10.4$~\pm~$0.1\\
		140209A		& 1.408		& 1.344     & 2.752		& `n9',`na',`nb',`b1'&9.0$~\pm~$0.1\\
		150819B		& 0.96		& $-$0.064    & 0.896		& `n2',`n9',`na',`b1'&8.1$~\pm~$0.1\\
		170206A		& 1.168		& 0.208     & 1.376		& `n9',`na',`nb',`b1'&10.2$~\pm~$0.2\\
		171108A		& 0.032		& $-$0.016    & 0.016		& `n9',`na',`nb',`b1'&10.4$~\pm~$0.6\\
		171126A		& 1.472		& 0       & 1.472		& `n0', `n1',`n2',`n3'&7.2$~\pm~$0.1\\
		180703B		& 1.536		& 0.128     & 1.664		& `n0','n1',`n3',`b0'&8.8$~\pm~$0.1\\
		181222B		& 0.576		& 0.032     & 0.608		& `n3',`n4',`n7',`b0'&36.2$~\pm~$0.1	\\
		\bottomrule
	\end{tabular}
\end{table}

\subsection{Method}
\subsubsection{Lightcurve Fitting}
GBM  {Time-Tagged Event (TTE, 2 \textmu s temporal resolution)} data of three NaI detectors  {is employed}, which were binned into 10 ms in our lightcurve fitting except for GRB 171108A, which takes 2 ms bins for its very short $T_{90}$ of 32 ms.  The energy band is selected ranging from 50 keV to 300 keV.
GRB lightcurves are usually irregular, but could be decomposed as several pulses, most of which are described as the fast-rising exponential-decay profile (FRED). FRED can be presented same as that in \citep{Paynter2021},
\begin{equation}
	S(t|A,T_{\rm start},T_{\rm rise},\xi) = A e^{  - \xi \left(
		\frac{t - T_{\rm start}}{T_{\rm rise}} + \frac{T_{\rm rise}}{t-T_{\rm start}}  -2 \right )  }
	\label{Eq:FRED}
\end{equation}
where $A$ is the amplitude, $T_{\rm start}$ is start time of the pulse and $T_{\rm rise}$ is the rise time interval before the peak, $\xi$ is an asymmetry parameter to represent the skewness of the FRED pulse. The decay time interval can be calculated by $T_{\rm decay} = \frac{1}{2}T_{\rm rise}\xi^{-1}[(1+4\xi)^{1/2}+1]$ \citep{Norris2005}. Therefore, we perform the fitting with several FRED pulses to nine GRBs in our sample, usually with 1, 2 or 3 FRED pulses which are named single-pulse SGRB, double-pulse SGRB and triple-pulse SGRB respectively.  {The maximum likelihood statistical method is employed in the LC fitting.}

\subsubsection{Spectral Energy Distribution Fitting}
GBM TTE data of all detectors in Table \ref{tab:info} are used in our spectral analysis. Instrument response files are selected with $rsp2$ files, we fit the background with an auto-selected orders polynomials using the Nelder-Mead method. Photons with energy ranging from 8 to 900 keV for NaIs are selected while from 200 keV to 40 MeV for BGOs.

We select 4 models to fit the gamma-ray spectra, e.g., the Band-function model (BAND), the cutoff power-law function model (CPL) and two blackbody function (BB)-included  models, such as BAND + BB and CPL + BB. In order to distinguish from the single BAND model and the single CPL model above, we named the BAND component and the BB component in the BAND + BB model while the CPL component and BB component in CPL + BB model.
BAND model is written as the so-called Band function  \citep{Band1993}, such as
\begin{eqnarray}
N(E)_{\rm BAND} = N_{\rm 0,BAND} \left\{ \begin{array}{ll}
(\frac{E}{E_{\rm piv}})^{\alpha} e^{[-E/E_{0}]}, & E\leq (\alpha-\beta)E_{0} \\
\\
(\frac{(\alpha-\beta)E_{0}}{E_{\rm piv}})^{(\alpha-\beta)} e^{(\beta-\alpha)}(\frac{E}{E_{\rm piv}})^{\beta}, & E\geq (\alpha-\beta)E_{0} \\
\end{array} \right.
	\label{Eq:Band}
\end{eqnarray}
where $\alpha$, $\beta$ is the photon index before and after the typical energy of $(\alpha- \beta)E_0$, and ${E_0}$ is the break energy in the $F_E (= E N_E)$ spectrum, note that the peak energy in the $E F_E (= E^2 N_E)$ spectrum $E_{\rm peak}= (2+\alpha) E_0$ \citep{Norris2005}. CPL could be regarded as the lower energy segment of the BAND model but with an exponential cutoff-power-law decay in the high-energy band, such as
\begin{equation}
	N_{E}({\rm CPL}) = N_{\rm 0,CPL} (\frac{E}{E_{\rm piv}})^{\Gamma} e^{-E/E_{\rm c}},
\end{equation}
where $\Gamma$ is the photon index and $E_{\rm c}$ is the cutoff energy. $E_{\rm piv}$ in both models is the pivot energy and fixed at 100 keV, which is most to adopt the observations.
BB component is usually modified by the standard Planck spectrum, which is given by the photon flux,
\begin{equation}
	N_{E}({\rm BB}) =  N_{\rm 0,BB}  \frac{E^2}{exp[E/kT]-1},
	\label{Eq:BB}
\end{equation}
where $k$ is the Boltzmann's constant, and the joint parameter $kT$ as a output parameter in common. The BB component is the additive spectral compoennt in our spectral analysis,  {such as BAND + BB} and CPL + BB. In all spectral models, $N_0$ is the normalisation.

For each spectral fitting, a likelihood value $L(\vec{\theta})$ as the function of the free parameters $\vec{\theta}$ is derived.  {The value} of the Bayesian Information Criterion (BIC; \citep{Schwarz1978}), defined as \mbox{BIC = $-$2ln$L(\vec{\theta})$+$k \ln n$,} are calculated, where $k$ is the number of free parameters to be estimated and $n$ is the number of observations (the sum of the selected GBM energy channels). In this work, the Multi-Mission Maximum Likelihood package (3ML; \citep{Vianello2015}) are employed to carry out all the spectral analysis and the parameter estimation, with the \textit{emcee} sampling method.

\section{Result \label{result}}

\subsection{Multiple  {Pulses}}
Results of the lightcurve fitting are presented in Table ~\ref{tab:para_lc}.
For the single-pulse GRBs, $T_{\rm rise}$ is about 0.03 s and 0.05 s for GRB 171108A and GRB 120323A respectively, while about 0.27 s and 0.59 s for GRB 171126A and GRB 140209A respectively.
For 4 GRBs with two FRED pulses (double-pulse SGRB), it is found that they have the similar pattern of the rising time, such as  the second rising time $T_{\rm rise, 2}$ is about 1 to  {3 times the first} rising time $T_{\rm rise, 1}$.
For the only GRB with three FRED pulses (triple-pulse SGRB), GRB 170206A has the most energetic flux in the sample. Its first two rising times are compared, e.g., 0.16 s, 0.12 s respectively. Its last rising  {time is 0.54 s}, which is longer than that in the former two pulses.

\begin{figure}[H]
	\subfigure[090227B]{
		\includegraphics[width=0.2\textwidth]{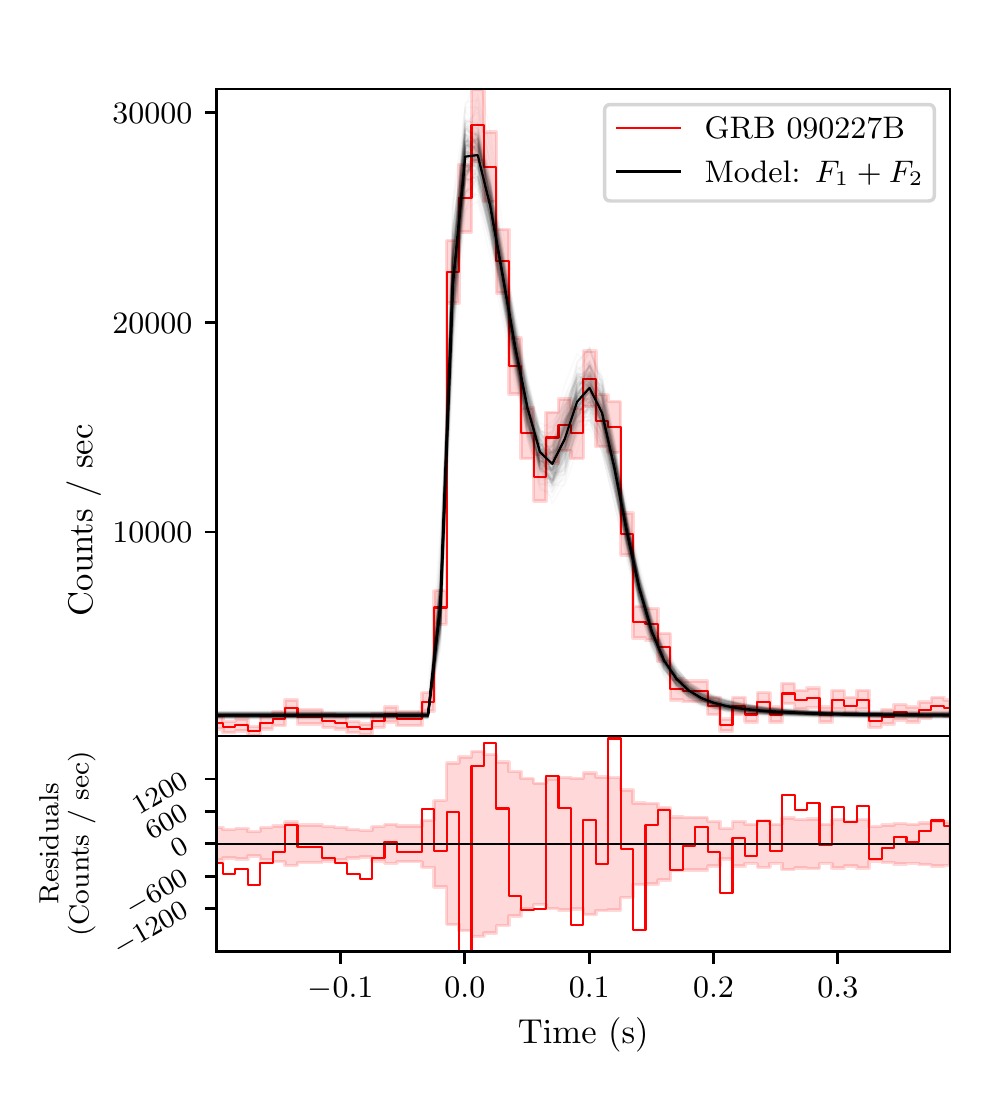}}
	\subfigure[120323A]{
		\includegraphics[width=0.2\textwidth]{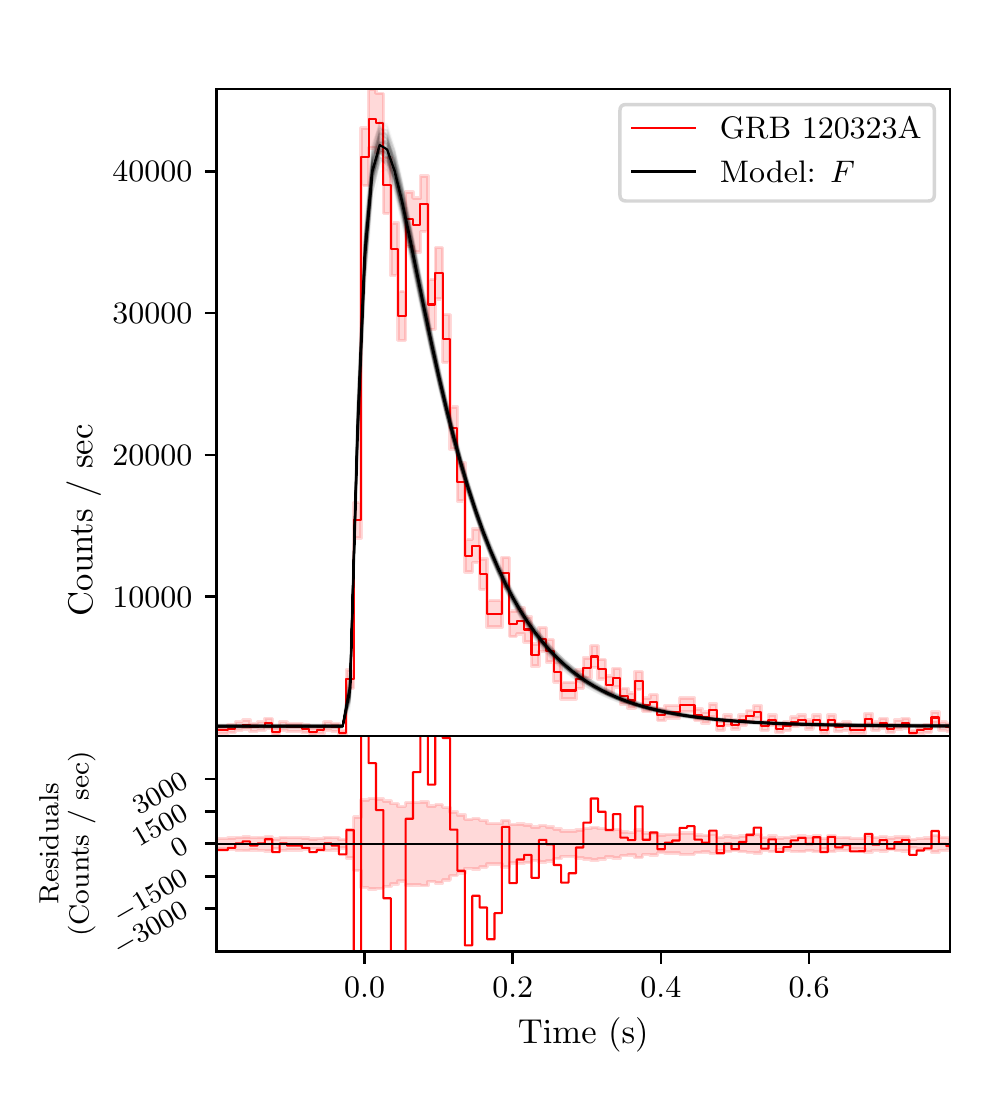}}
	\subfigure[140209A]{
		\includegraphics[width=0.2\textwidth]{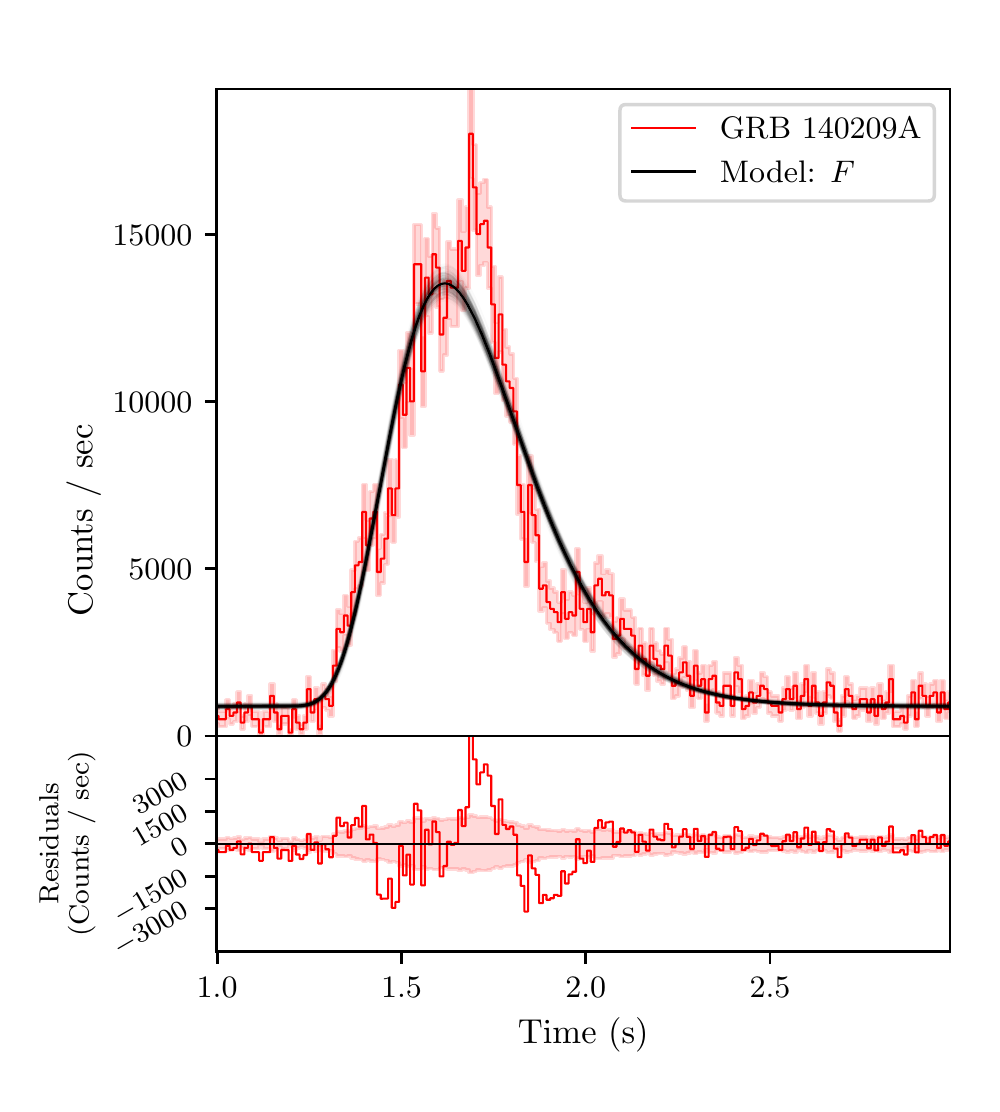}}
	\subfigure[150819B]{
		\includegraphics[width=0.2\textwidth]{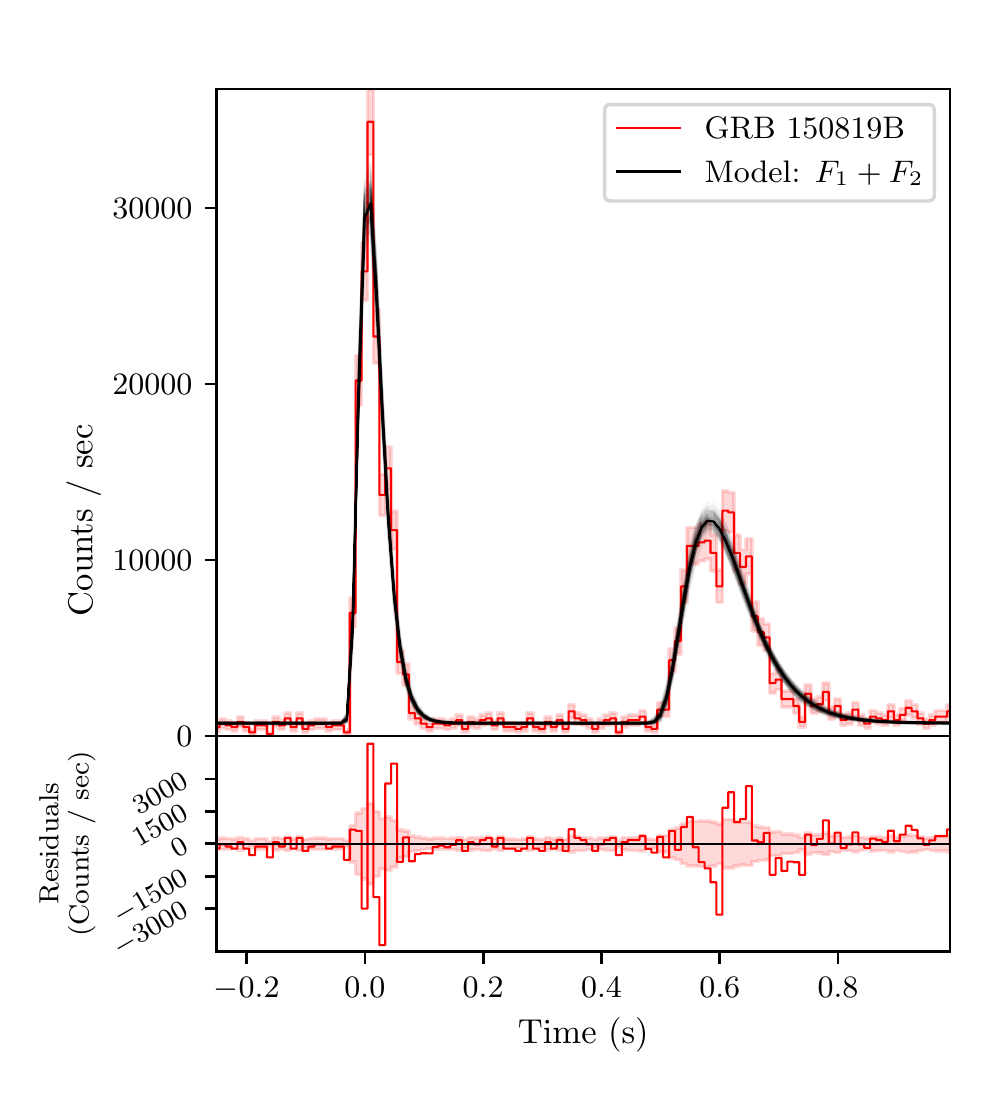}}
	\subfigure[170206A]{
		\includegraphics[width=0.2\textwidth]{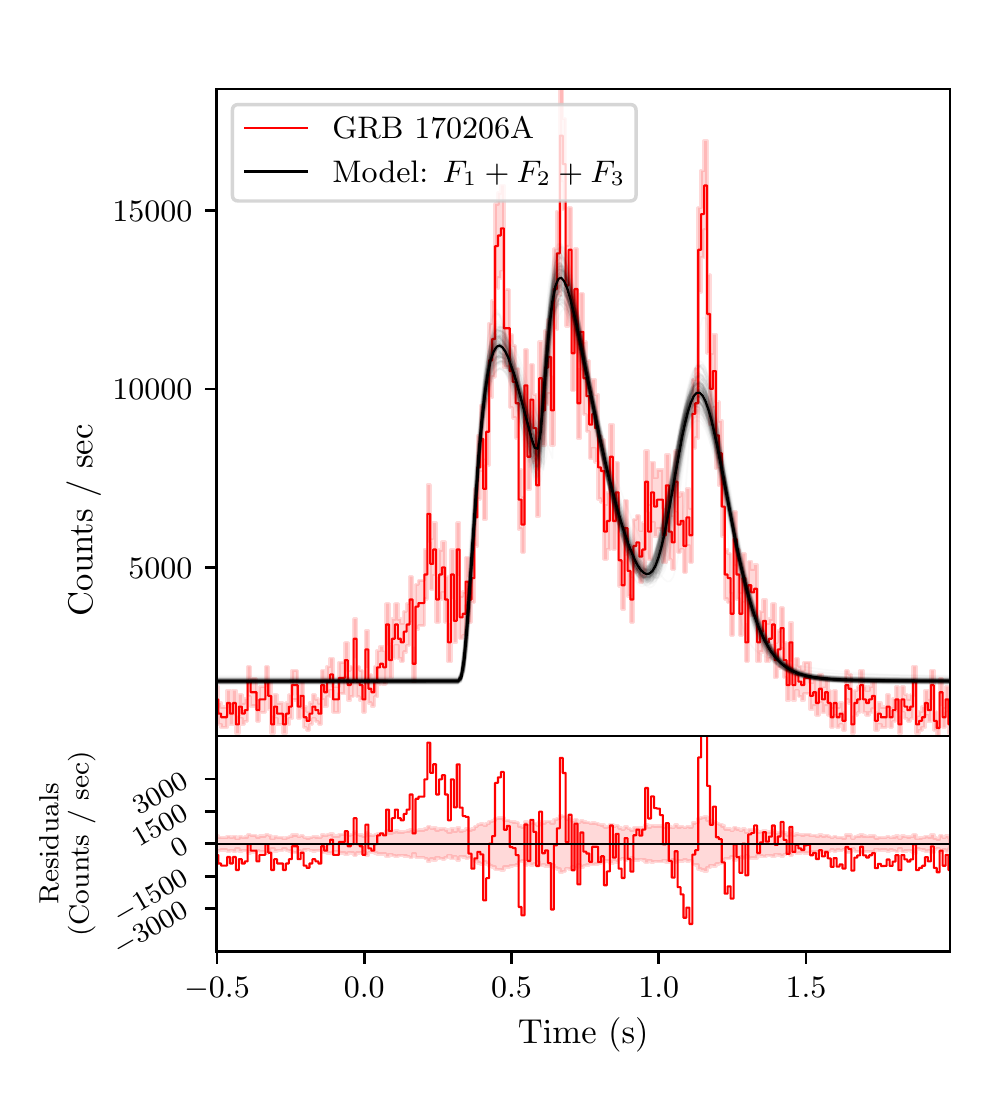}}
	\subfigure[171108A]{
		\includegraphics[width=0.2\textwidth]{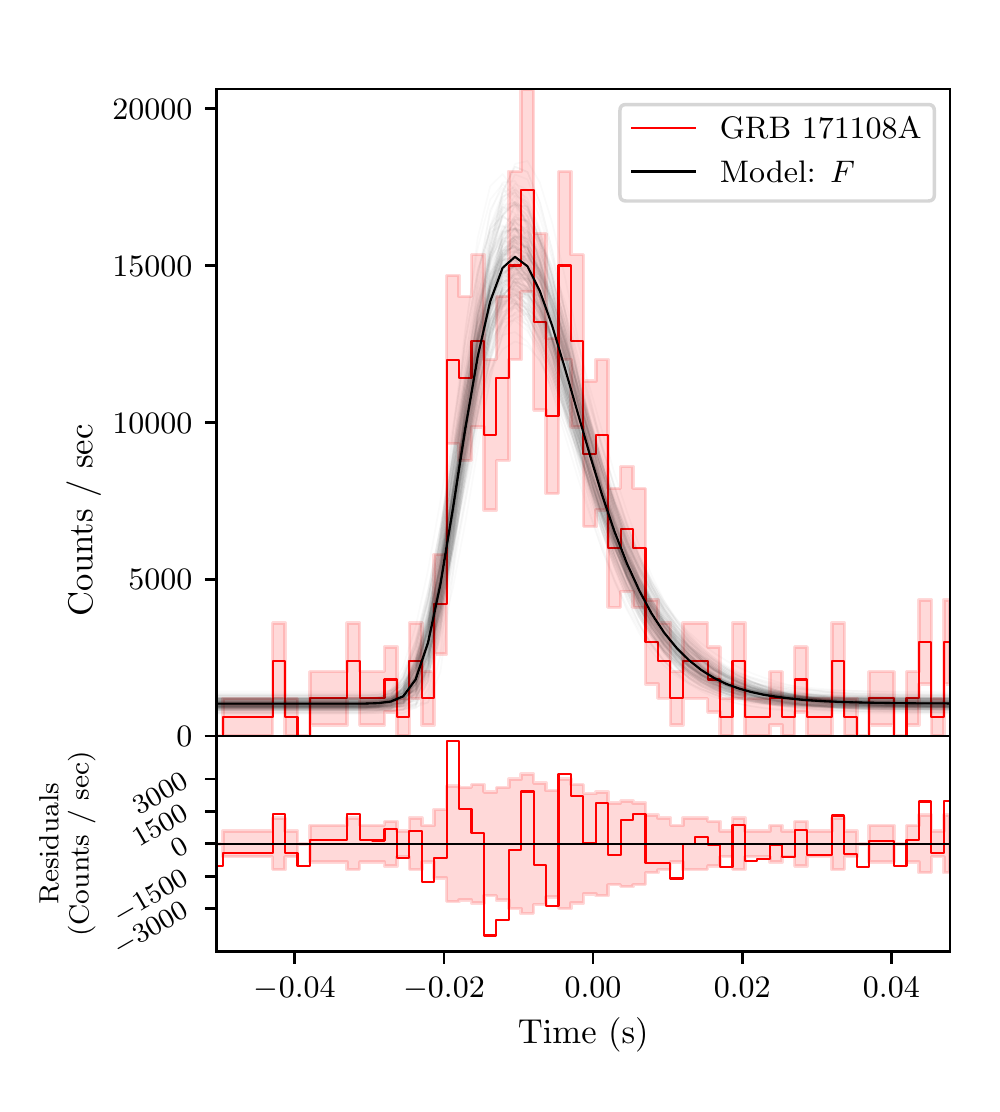}}
	\subfigure[171126A]{
		\includegraphics[width=0.2\textwidth]{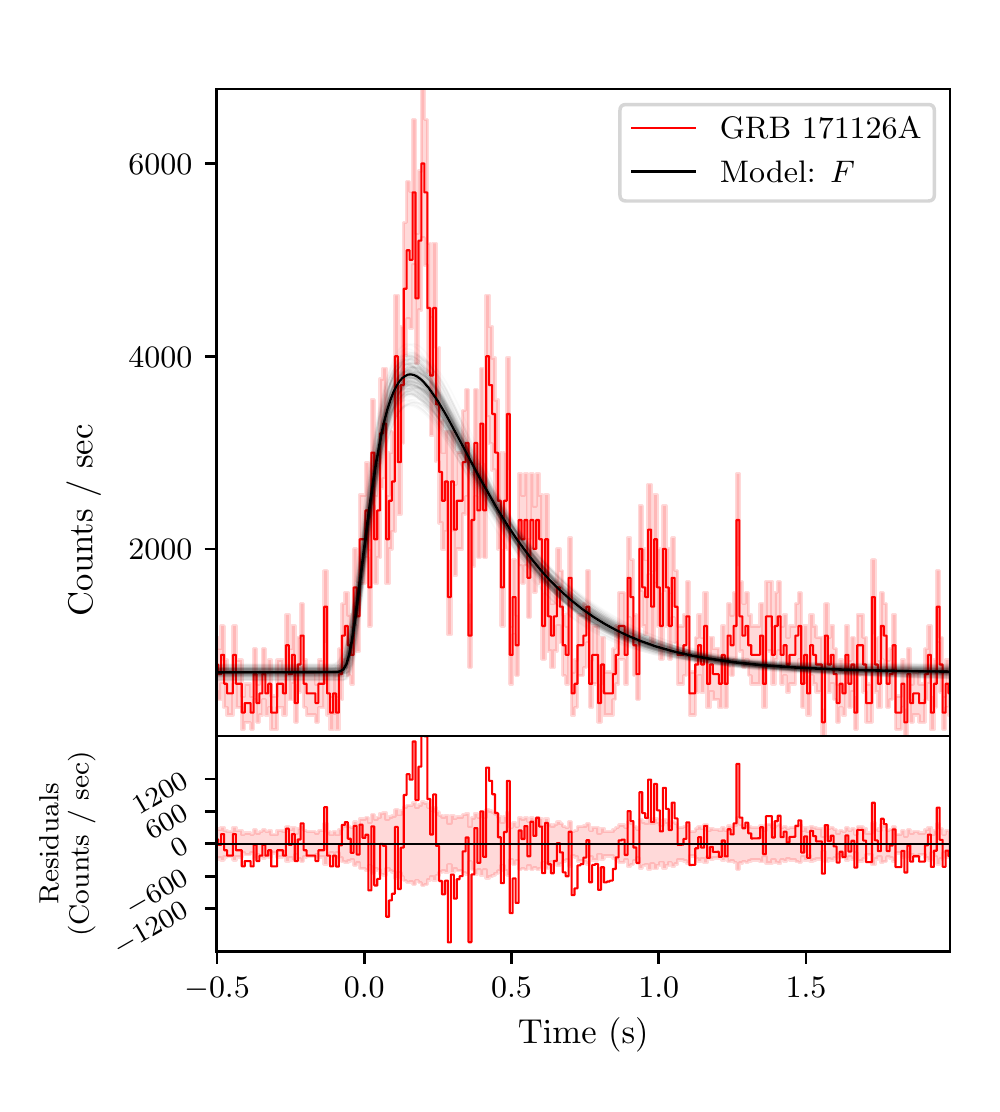}}
	\subfigure[180703B]{
		\includegraphics[width=0.2\textwidth]{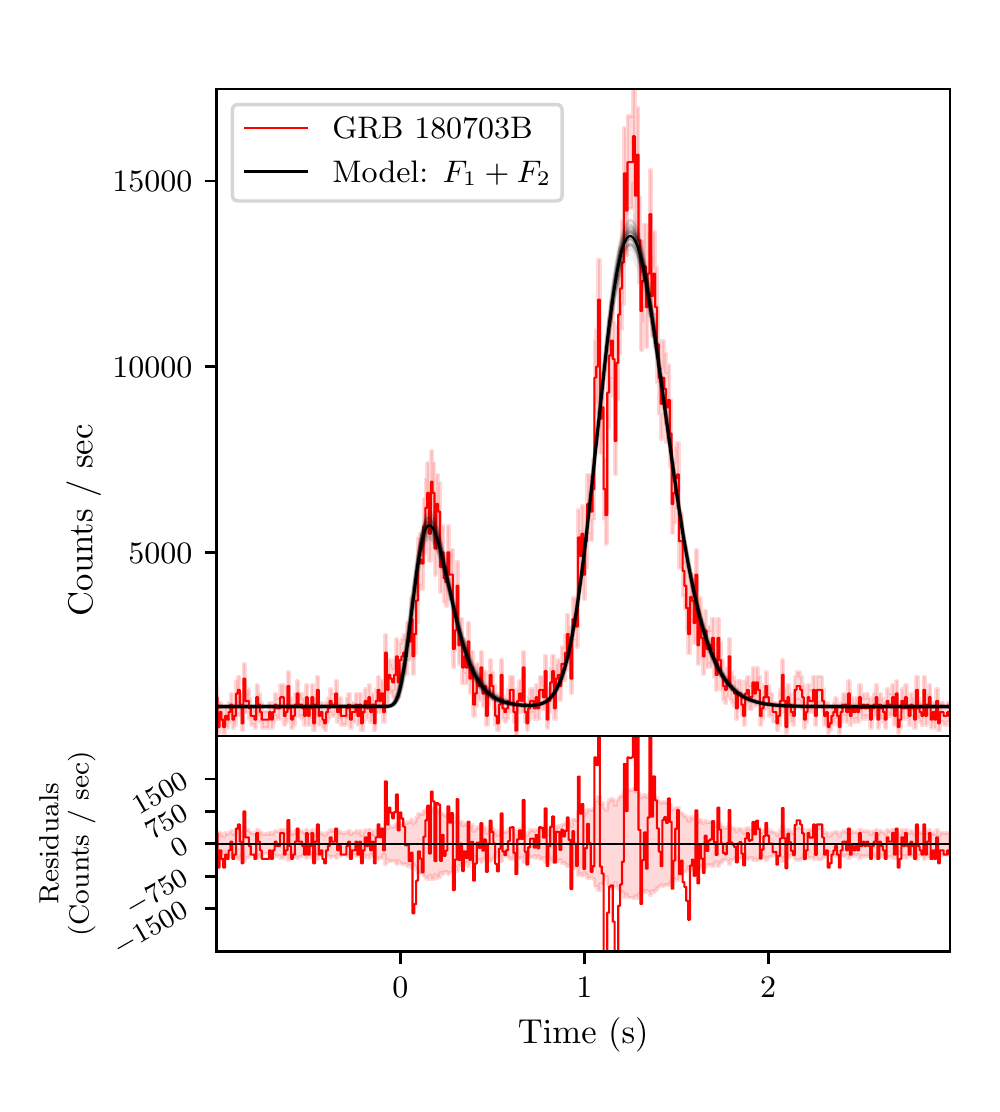}}
	\subfigure[181222B]{
		\includegraphics[width=0.2\textwidth]{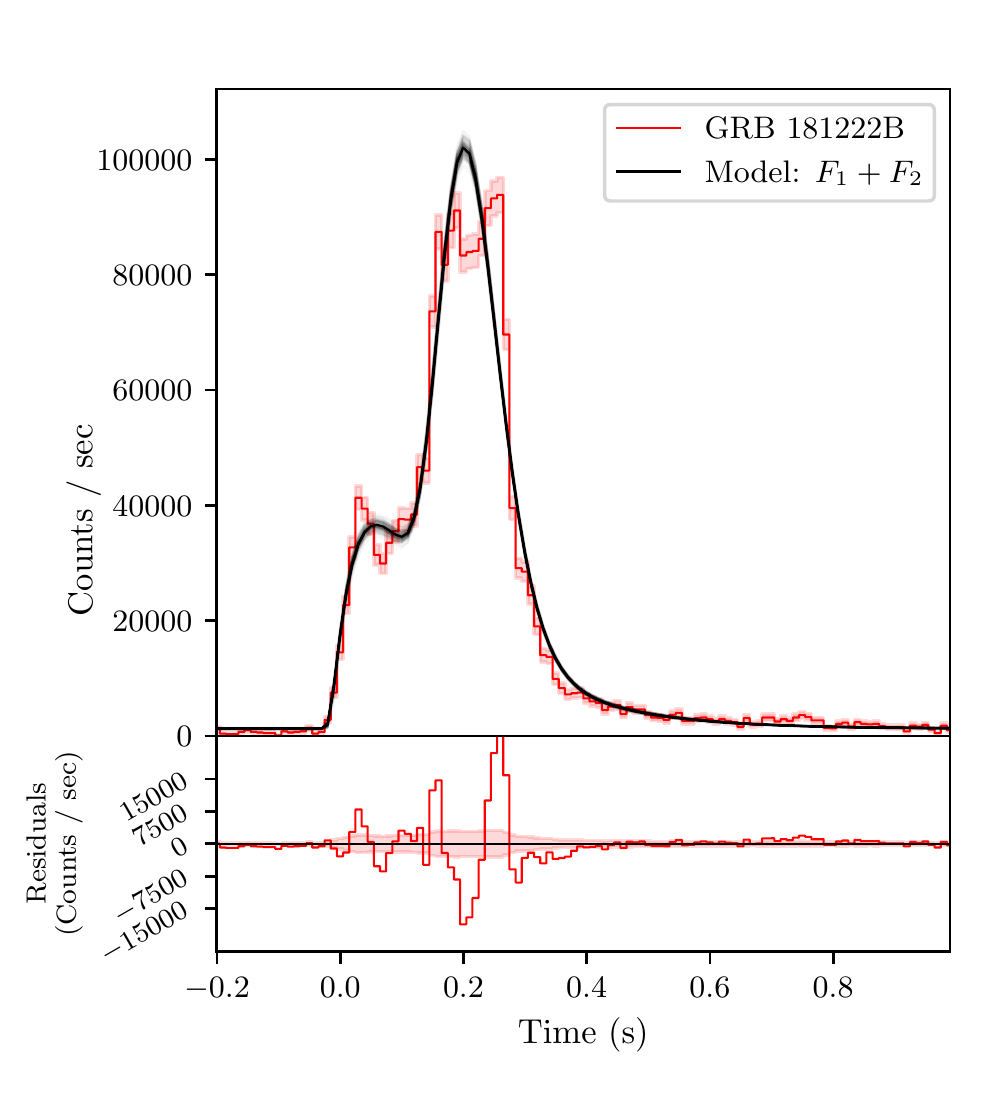}}
	\caption{Observaional lightcurves in the energy band from 50 keV to 300 keV and the best-fit lines modelled by the FRED functions. For top pannel of each GRB, the red histogram represents \mbox{50--300~keV} observational lightcurve, and solid black line is the best fit with the empirical function (FRED). The bottom pannel in ecah GRB is the residual between the observation and the fitting model.}
	\label{fig:lightcurve}
\end{figure}

Figure ~\ref{fig:lightcurve} shows the observational lightcurves and the fitting curves of 9 GRBs in our sample. The red histograms represent lightcurves, and solid black lines are the best fittings by FRED profiles.

\begin{table}[H]
	\tiny
	\caption{Derived parameters of GRB FRED pulses \label{tab:para_lc}.}
		\begin{tabular}{m{1.8cm}<{\centering}ccccccccc}
			\toprule
			\multirow{3}{0.5cm}{\textbf{GRB} 
}&\multicolumn{9}{c}{\textbf{Models}}\\
\cmidrule{2-10}
&\multicolumn{3}{c}{\boldmath{$Pulse_{1}$}}&\multicolumn{3}{c}{\boldmath{$Pulse_{2}$}}&\multicolumn{3}{c}{\boldmath{$Pulse_{3}$}}\\
\cmidrule{2-10}
&\boldmath{$T_{\rm start,1}$}&\boldmath{$T_{\rm rise,1}$}&\boldmath{$FWHM_{1}$}&\boldmath{$T_{\rm start,2}$}&\boldmath{$T_{\rm rise,2}$}&\boldmath{$FWHM_{2}$}&\boldmath{$T_{\rm start,3}$}&\boldmath{$T_{\rm rise,3}$}&\boldmath{$FWHM_{3}$}\\
&\boldmath{(s)}&\boldmath{(s)}&\boldmath{(s)}&\boldmath{(s)}&\boldmath{(s)}&\boldmath{(s)}&\boldmath{(s)}&\boldmath{(s)}&\boldmath{(s)}\\
			\midrule		
			 090227B&${-0.03}\pm{0.03}$&${0.04}\pm{0.01}$&${0.06}\pm{0.02}$&${-0.02}\pm{0.02}$&${0.07}\pm{0.04}$&${0.05}\pm{0.04}$&-&-&-\\
			120323A&${-0.03}\pm{0.01}$&${0.05}\pm{0.01}$&${0.13}\pm{0.01}$&-&-&-&-&-&-\\
			140209A&${1.03}\pm{0.02}$&${0.59}\pm{0.03}$&${0.44}\pm{0.03}$&-&-&-&-&-&-\\
			 150819B&${-0.05}\pm{0.01}$&${0.05}\pm{0.01}$&${0.04}\pm{0.01}$&${0.45}\pm{0.01}$&${0.14}\pm{0.02}$&${0.13}\pm{0.02}$&-&-&-\\
			 170206A&${0.30}\pm{0.01}$&${0.16}\pm{0.01}$&${0.27}\pm{0.03}$&${0.57}\pm{0.02}$&${0.12}\pm{0.02}$&${0.24}\pm{0.06}$&${0.60}\pm{0.03}$&${0.54}\pm{0.03}$&${0.22}\pm{0.02}$\\
			171108A&${-0.04}\pm{0.01}$&${0.03}\pm{0.01}$&${0.02}\pm{0.01}$&-&-&-&-&-&-\\
			171126A&${-0.11}\pm{0.02}$&${0.27}\pm{0.03}$&${0.48}\pm{0.08}$&-&-&-&-&-&-\\
			 180703B&${-0.19}\pm{0.01}$&${0.35}\pm{0.01}$&${0.24}\pm{0.01}$&${0.70}\pm{0.01}$&${0.51}\pm{0.01}$&${0.42}\pm{0.01}$&-&-&-\\
			 181222B&${-0.03}\pm{0.01}$&${0.09}\pm{0.01}$&${0.21}\pm{0.01}$&${-0.03}\pm{0.01}$&${0.23}\pm{0.01}$&${0.11}\pm{0.01}$&-&-&-\\
			\bottomrule
		\end{tabular}
\end{table}

\vspace{-20pt}

In order to analyze the relationship between the rising time ($T_{\rm rise}$) phase and the whole pulse time width, full time with in half maximum (FWHM) is measured, which  {is also presented in Table~\ref{tab:para_lc}}.
Then we test the correlation between FWHM and $T_{\rm rise}$ by the linear fitting in logarithmic space, such as
\begin{equation}
	log ({\rm FWHM}) = m + n log(T_{\rm rise})
\end{equation}
where $m$ and $n$ are the free parameters.
This fitting is performed by the basic linear regression analysis in the popular $Origin$ scientific package, which can give the coefficient of determination $R^2$ ($0 < R^2 <1$). For the linear fit, two variables, such as FWHM and $T_{\rm rise}$ in this section, are positively correlated if the Pearson-correlation coefficient $R$ ($-1 < R <1$) is close to $1$. For example, a strong correlation can be claimed when $R$ $>$ 0.8 while a
moderate correlation can be claimed when 0.5 $<R<$ 0.8~\cite{Newton1999}. We also calculate the probability $p$ of the null hypothesis, which can be described as the confidence level of 1$-p$ for the correlation.

\begin{figure}[H]
	\includegraphics[width=0.60\textwidth]{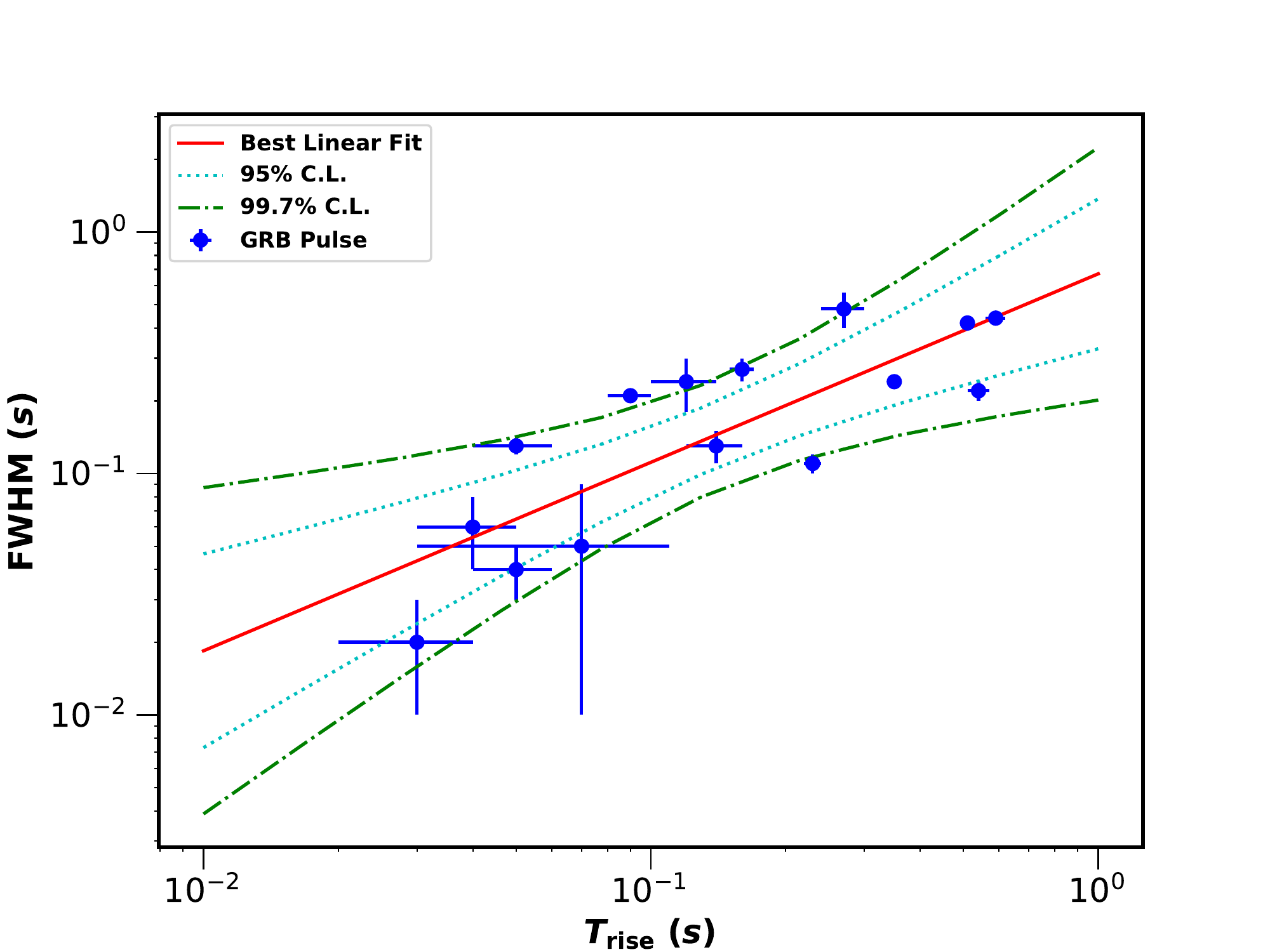}
	\caption{Correlation of the rising time ($T_{\rm rise}$) and the full width at half maxima (FWHM) of individual pulses in all selected short GRBs. 95\% and 99.7\% confidence leverls are plotted by light blue dotted line and green dashed line respectively.}
	\label{fig:time}
\end{figure}

For the individual pulse in our sample, we found that $T_{\rm rise}$ and FWHM are strongly correlated with each other, which is plotted in Figure ~\ref{fig:time}, with $R$ = 0.82, $m$ = $-0.17\pm0.14$ and $n$ = $0.78\pm0.15$.
The best fit for the correlation is written as
\begin{equation}
	log ({\rm FWHM}) = (-0.17\pm0.14) + (0.78\pm0.15) log(T_{\rm rise}),
	\label{eq:correlation1}
\end{equation}
$p$ is about 1.9$\times$10$^{-4}$,  {which also favors a strong positive correlation}.
This strong correlaiton implies those pulses are emitted within one impulsive explosion.

\vspace{-6pt}

\subsection{Multiple Spectral Components}
\begin{figure}[H]
	\subfigure[]{
		\includegraphics[width=0.2\textwidth]{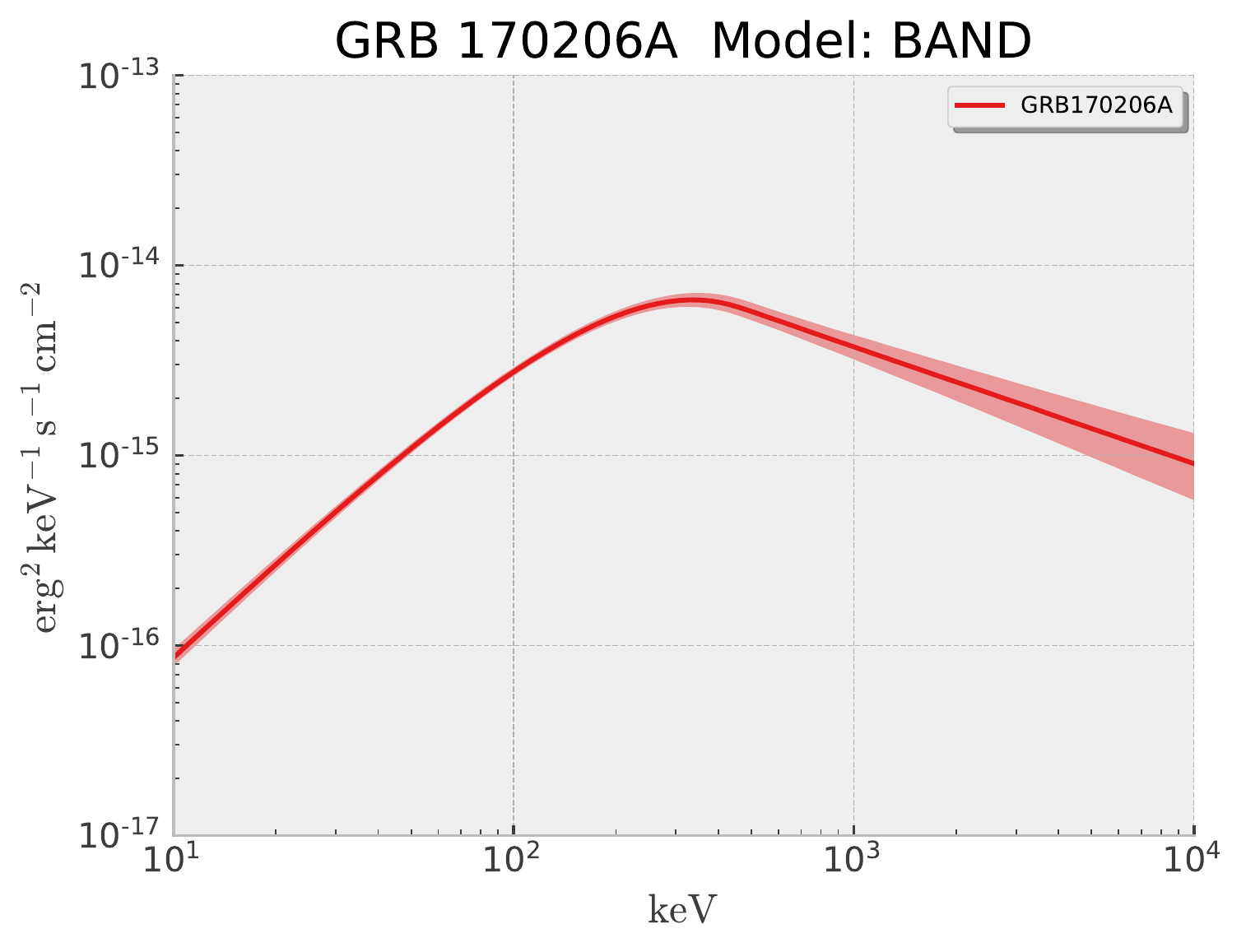}}
	\subfigure[]{
		\includegraphics[width=0.2\textwidth]{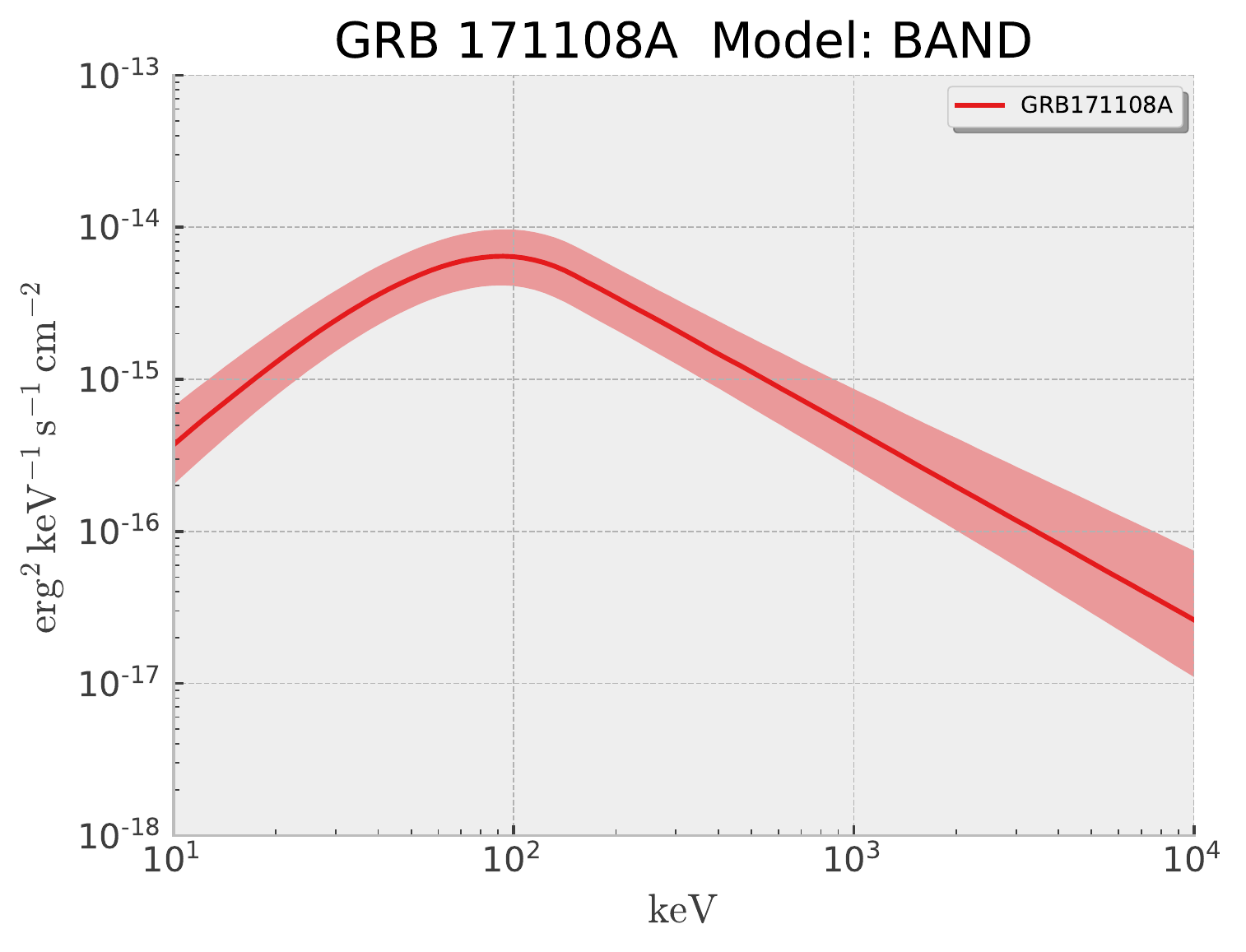}}
	\subfigure[]{
		\includegraphics[width=0.2\textwidth]{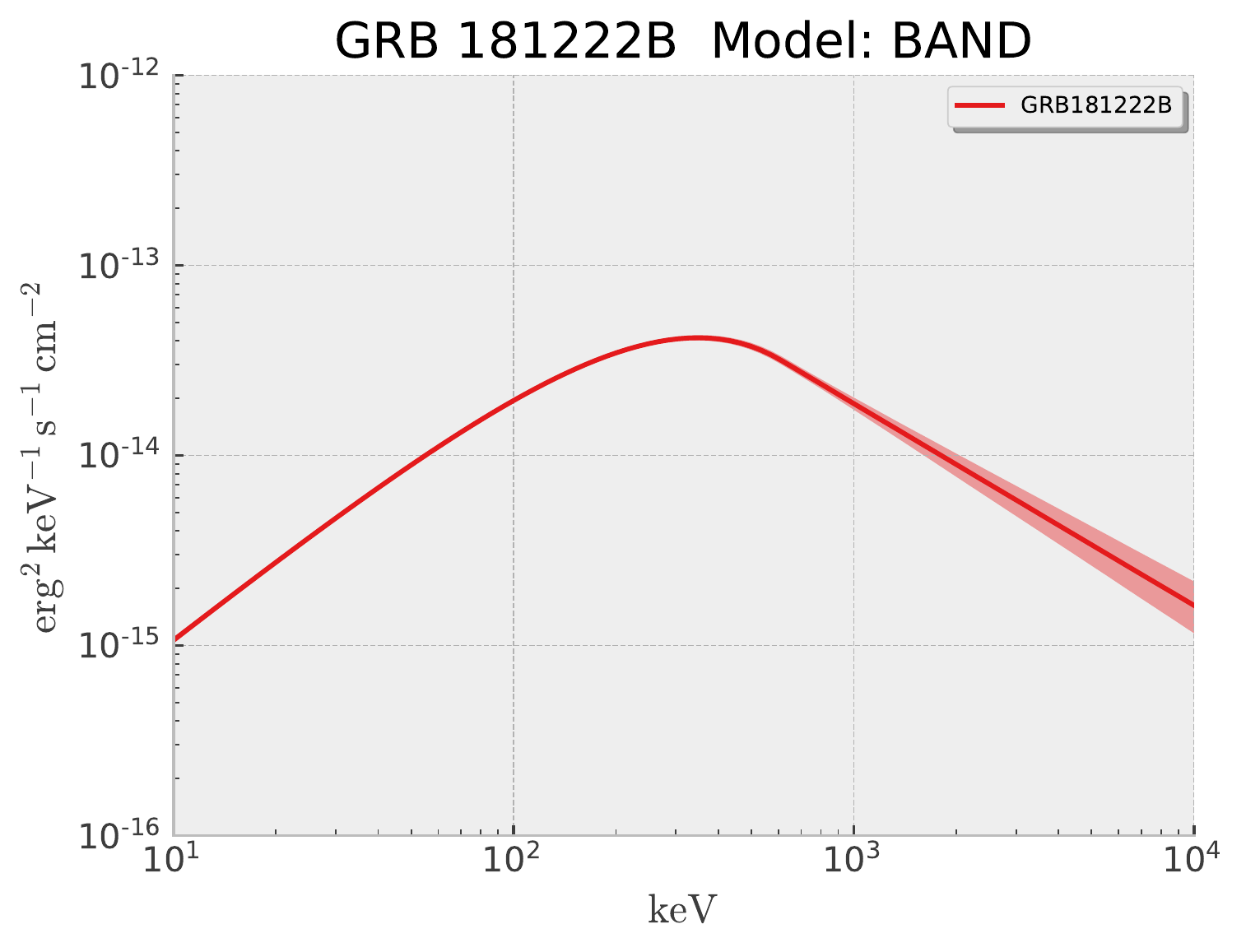}}
	\subfigure[]{
		\includegraphics[width=0.2\textwidth]{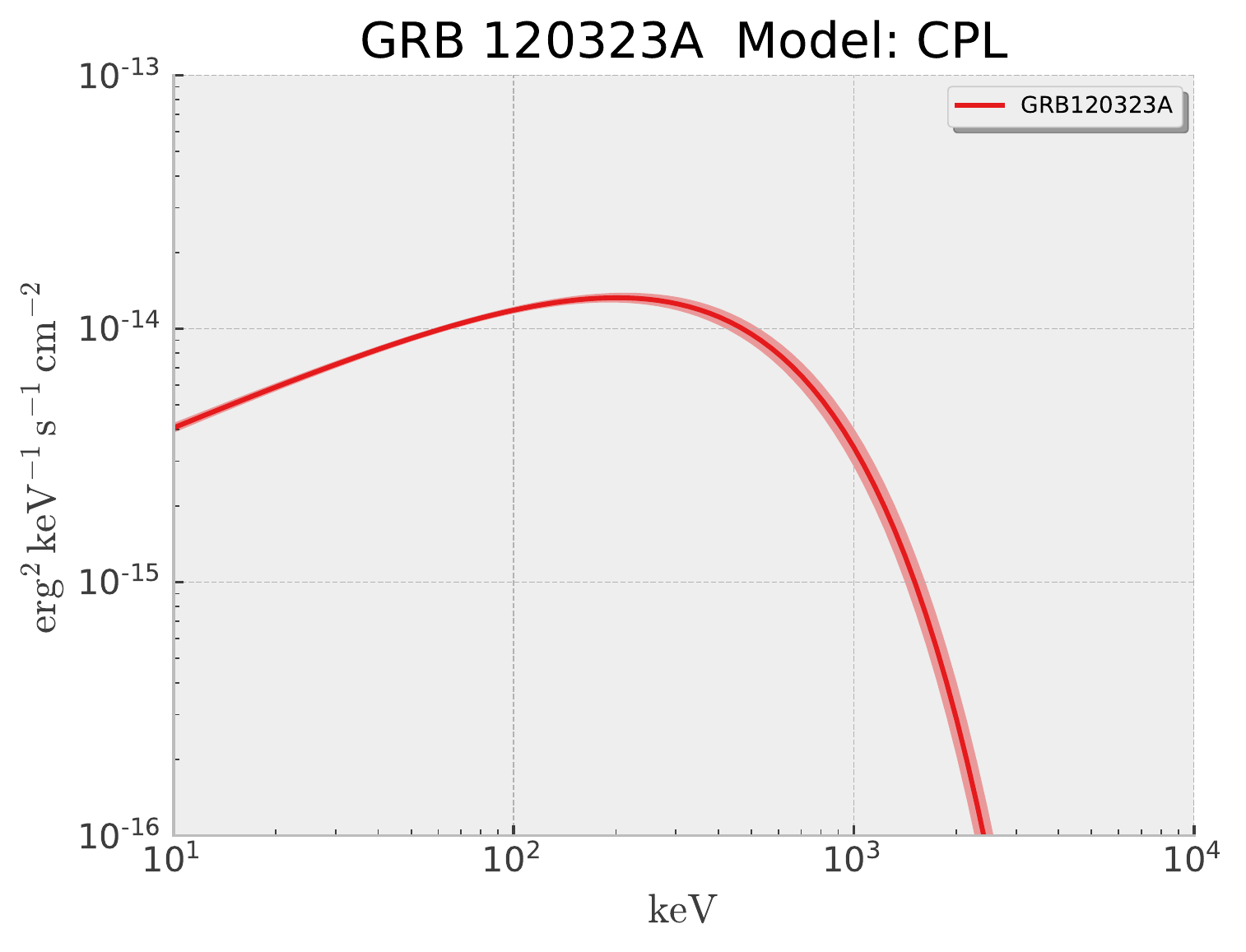}}
	\subfigure[]{
		\includegraphics[width=0.2\textwidth]{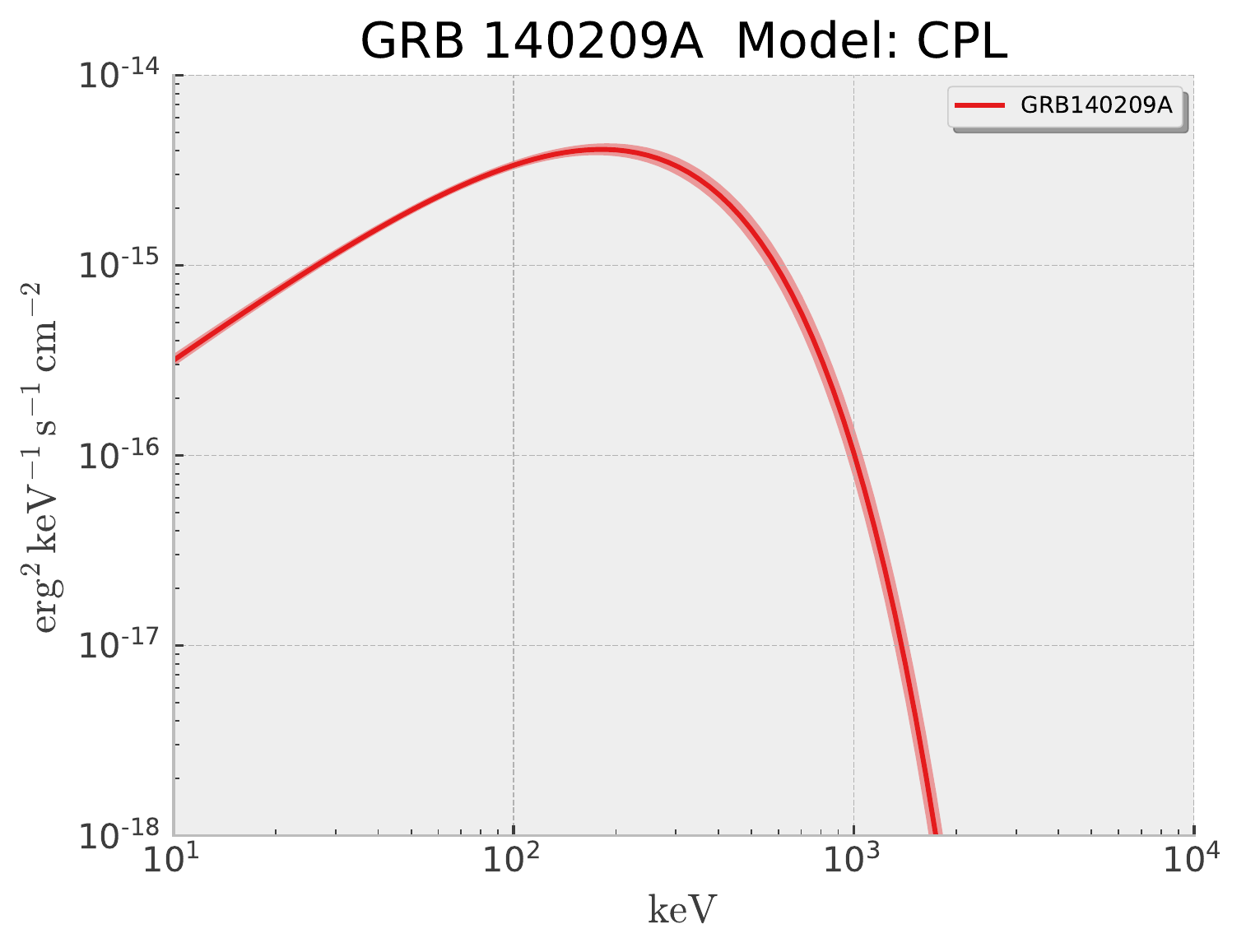}}
	\subfigure[]{
		\includegraphics[width=0.2\textwidth]{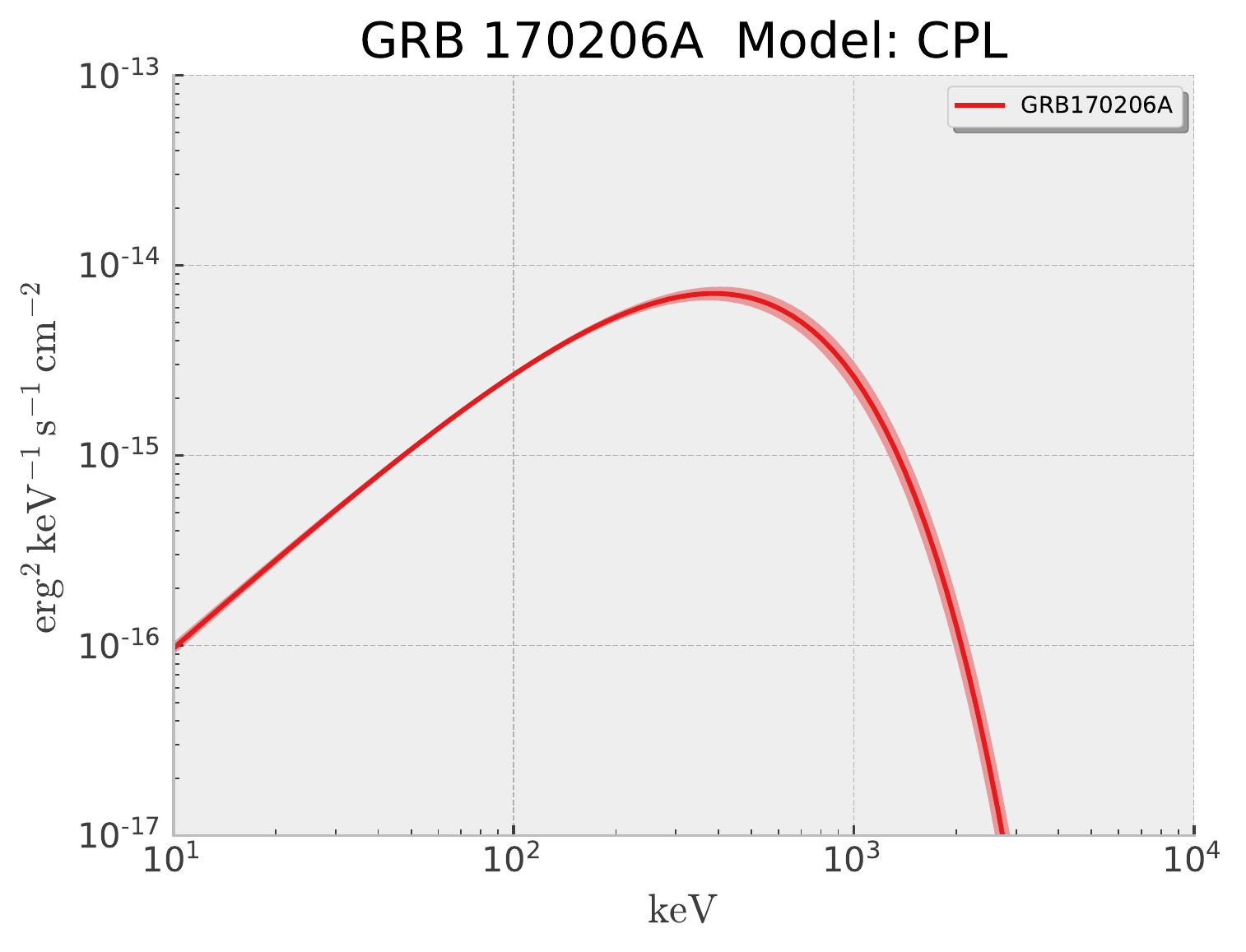}}
	\subfigure[]{
		\includegraphics[width=0.2\textwidth]{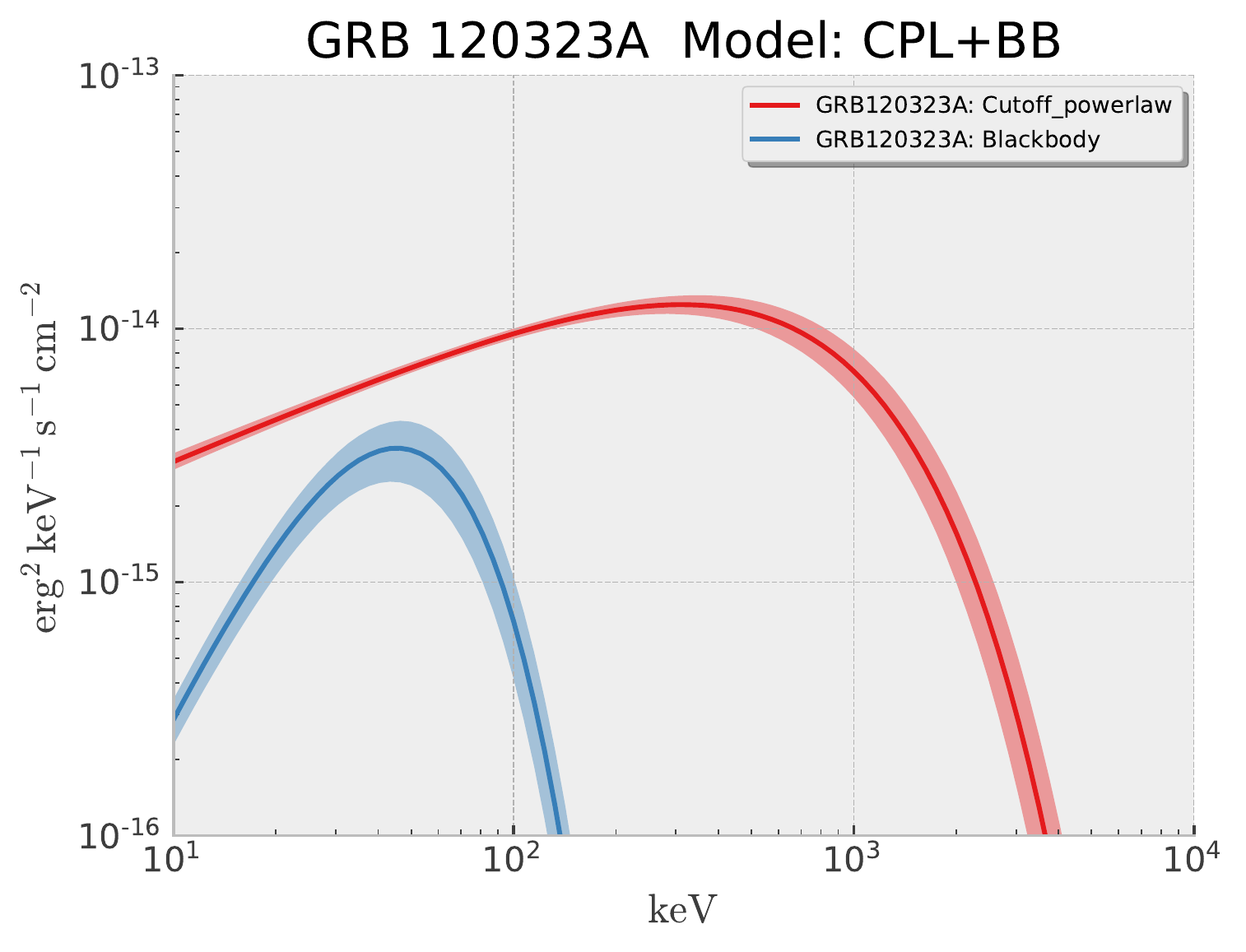}}
	\subfigure[]{
		\includegraphics[width=0.2\textwidth]{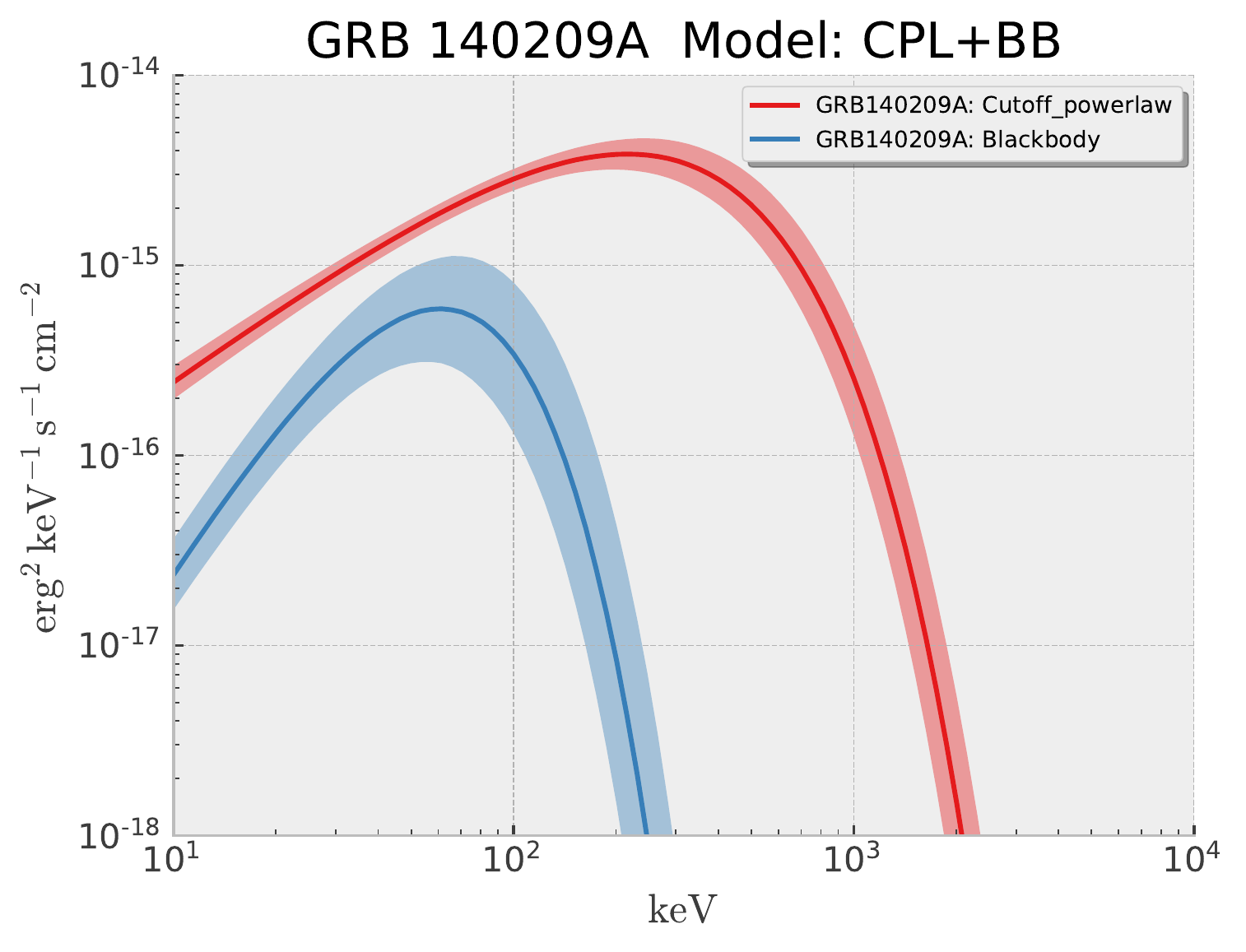}}
	\subfigure[]{
		\includegraphics[width=0.2\textwidth]{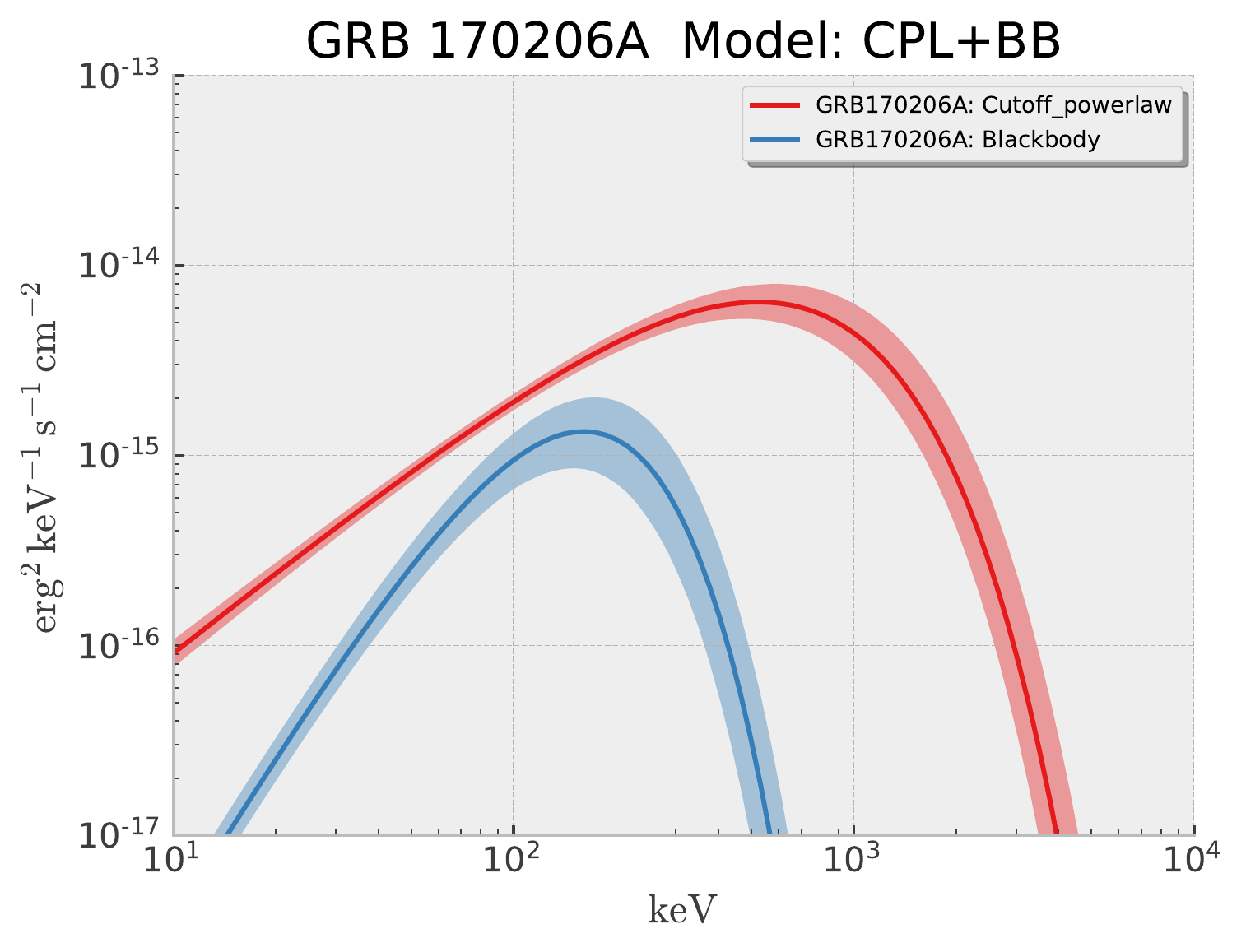}}
	\caption{Examples for the SED fitting by four models. (a) + (b) + (c): BAND; (d) + (e) + (f): CPL; (g) + (h) + (i): CPL + BB.}
	\label{fig:sed}
\end{figure}

We obtained the resultant spectral parameters of each GRB as well as the BIC values,  {presented} in Table ~\ref{tab:sed}. Firstly, BIC value of one model is enough smaller than that in other mdoels, i.e., $\Delta$BIC $>6$,  {is selected} as the best-fit model~\cite{Kass1995}. If two or more models have the $\Delta$BIC $<6$ with each other, then we named them the compareable models. Secondly, we will reject a model if the resultant parameters in a candidate model  {are unreasonable}, such as $kT$ is less than 8 keV, which is the minimum photon energy we selected. For each model, we plot three GRBs as the example in Figure ~\ref{fig:sed}.
\begin{itemize}
	\item For GRB 090227B, CPL, BAND + BB and CPL + BB models are the compared models.
	
	\item For GRB 120323A, CPL + BB model is the best-fit model.

	\item For GRB 140209A, BAND and BAND + BB models are the compared models.

	\item For GRB 150819B, CPL and CPL + BB models are the compared models.

	\item For GRB 170206A, CPL + BB model is the best-fit model.

	\item For GRB 171108A, BAND, CPL and CPL + BB models are the compared models.

	\item For GRB 171126A, BAND and CPL models are the compared models.

	\item For GRB 180703B, BAND, CPL and CPL + BB models are the comared models.

	\item For GRB 181222B, BAND model is the best model.
\end{itemize}

Amongst four candidate models, the CPL + BB model is best model in two GRBs, such as GRB 120323A and GRB 170206A, and is the compared models in GRBs 090227B, 150819B, 171108A, 180703B.
For other 3 GRBs, although the CPL + BB model is not the best/compared model, the resultant $kT$ is well constrained. In order to test the possible correlation between the multiple spectral components, we thus choose the spectral fittings that are modelled by CPL + BB model in the following analysis.

\begin{table}[H]
	\caption{Resultant parameters by the spectral fitting.\label{tab:sed}}
	\begin{tabular}{m{1.4cm}<{\centering}cccccc}
		\toprule
		\boldmath{$GRB$}&\boldmath{$Model$}&\boldmath{$BIC$}&\boldmath{$\alpha / \Gamma$}&\boldmath{$E_{peak}/E_{c}$}&\boldmath{$\beta$}&\boldmath{${kT}$}\\
		&&&& \textbf{(keV) }&& \textbf{(keV)}\\
		\midrule
		\multirow{4}*{090227B}&BAND&1692&${-0.12}\pm{0.03}$&${1000}\pm{4}$&${-2.10}\pm{0.05}$&-\\
		\cmidrule{2-7}
		&CPL&1517&${-0.46}\pm{0.02}$&${1400}\pm{66}$&-&-\\
		\cmidrule{2-7}
		&BAND + BB&1522&${-0.42}\pm{0.04}$&${950}\pm{58}$&${-2.20}\pm{0.10}$&${490}\pm{29}$\\
		\cmidrule{2-7}
		&CPL + BB&1522&${-0.55}\pm{0.06}$&${1600}\pm{180}$&-&${270}\pm{66}$\\
		\midrule
		\multirow{4}*{120323A}&BAND&2151&${-0.84}\pm{0.07}$&${72}\pm{5}$&${-2.10}\pm{0.02}$&-\\
		\cmidrule{2-7}
		&CPL&2310&${-1.40}\pm{0.02}$&${350}\pm{21}$&-&-\\
		\cmidrule{2-7}
		&BAND  +  BB&2158&${-0.81}\pm{0.11}$&${71}\pm{7}$&${-2.10}\pm{0.02}$&${1}\pm{1}$\\
		\cmidrule{2-7}
		&CPL  +  BB&2097&${-1.40}\pm{0.03}$&${530}\pm{57}$&-&${12}\pm{1}$\\
		\midrule
		\multirow{4}*{140209A}&BAND&3194&${-0.52}\pm{0.05}$&${150}\pm{6}$&${-2.40}\pm{0.08}$&-\\
		\cmidrule{2-7}
		&CPL&3248&${-0.69}\pm{0.03}$&${140}\pm{6}$&-&-\\
		\cmidrule{2-7}
		&BAND + BB&3191&${-0.33}\pm{0.15}$&${160}\pm{11}$&${-2.50}\pm{0.09}$&${10}\pm{1}$\\
		\cmidrule{2-7}
		&CPL + BB&3220&${-0.70}\pm{0.07}$&${170}\pm{18}$&-&${16}\pm{2}$\\
		\midrule
		\multirow{4}*{150819B}&BAND&2670&${-1.10}\pm{0.03}$&${540}\pm{38}$&${-3.30}\pm{0.21}$&-\\
		\cmidrule{2-7}
		&CPL&2663&${-1.10}\pm{0.03}$&${600}\pm{55}$&-&-\\
		\cmidrule{2-7}
		&BAND + BB&2672&${-1.00}\pm{0.12}$&${510}\pm{66}$&${-3.20}\pm{0.24}$&${8}\pm{2}$\\
		\cmidrule{2-7}
		&CPL + BB&2665&${-1.10}\pm{0.11}$&${570}\pm{96}$&-&${6}\pm{6}$\\
		\midrule
		\multirow{4}*{170206A}&BAND&3056&${-0.30}\pm{0.04}$&${340}\pm{14}$&${-2.6}\pm{0.13}$&-\\
		\cmidrule{2-7}
		&CPL&3054&${-0.40}\pm{0.03}$&${240}\pm{11}$&-&-\\
		\cmidrule{2-7}
		&BAND + BB&3062&${-0.54}\pm{0.14}$&${430}\pm{85}$&${-3.00}\pm{0.39}$&${38}\pm{9}$\\
		\cmidrule{2-7}
		&CPL + BB&3048&${-0.57}\pm{0.06}$&${370}\pm{44}$&-&${41}\pm{4}$\\
		\midrule
		\multirow{4}*{171108A}&BAND&6&${0.15}\pm{0.20}$&${92}\pm{6}$&${-3.30}\pm{0.21}$&-\\
		\cmidrule{2-7}
		&CPL&2&${-0.03}\pm{0.19}$&${51}\pm{7}$&-&-\\
		\cmidrule{2-7}
		&BAND + BB&12&${0.49}\pm{0.61}$&${100}\pm{23}$&${-3.30}\pm{0.26}$&${10}\pm{10}$\\
		\cmidrule{2-7}
		&CPL + BB&7&${0.32}\pm{0.44}$&${45}\pm{10}$&-&${8}\pm{5}$\\
		\midrule
		\multirow{4}*{171126A}&BAND&2988&${-0.38}\pm{0.08}$&${88}\pm{4}$&${-3.00}\pm{0.23}$&-\\
		\cmidrule{2-7}
		&CPL&2989&${-0.49}\pm{0.07}$&${62}\pm{5}$&-&-\\
		\cmidrule{2-7}
		&BAND + BB&2999&${-0.37}\pm{0.12}$&${90}\pm{5}$&${-3.10}\pm{0.23}$&${1}\pm{1}$\\
		\cmidrule{2-7}
		&CPL + BB&2995&${-0.49}\pm{0.11}$&${65}\pm{9}$&-&${8}\pm{6}$\\
		\midrule
		\multirow{4}*{180703B}&BAND&3482&${-0.65}\pm{0.04}$&${140}\pm{5}$&${-3.10}\pm{0.22}$&-\\
		\cmidrule{2-7}
		&CPL&3481&${-0.70}\pm{0.03}$&${110}\pm{4}$&-&-\\
		\cmidrule{2-7}
		&BAND + BB&3487&${-0.63}\pm{0.06}$&${140}\pm{10}$&${-3.10}\pm{0.28}$&${8}\pm{7}$\\
		\cmidrule{2-7}
		&CPL + BB&3483&${-0.67}\pm{0.06}$&${120}\pm{8}$&-&${12}\pm{2}$\\
		\midrule
		\multirow{4}*{181222B}&BAND&2677&${-0.58}\pm{0.01}$&${350}\pm{6}$&${-3.10}\pm{0.11}$&-\\
		\cmidrule{2-7}
		&CPL&2719&${-0.61}\pm{0.01}$&${270}\pm{6}$&-&-\\
		\cmidrule{2-7}
		&BAND + BB&2687&${-0.58}\pm{0.02}$&${350}\pm{7}$&${-3.10}\pm{0.10}$&${1}\pm{1}$\\
		\cmidrule{2-7}
		&CPL + BB&2721&${-0.66}\pm{0.04}$&${290}\pm{20}$&-&${40}\pm{16}$\\
		\bottomrule
	\end{tabular}
\end{table}

\begin{figure}[H]
	\includegraphics[width=0.6\textwidth]{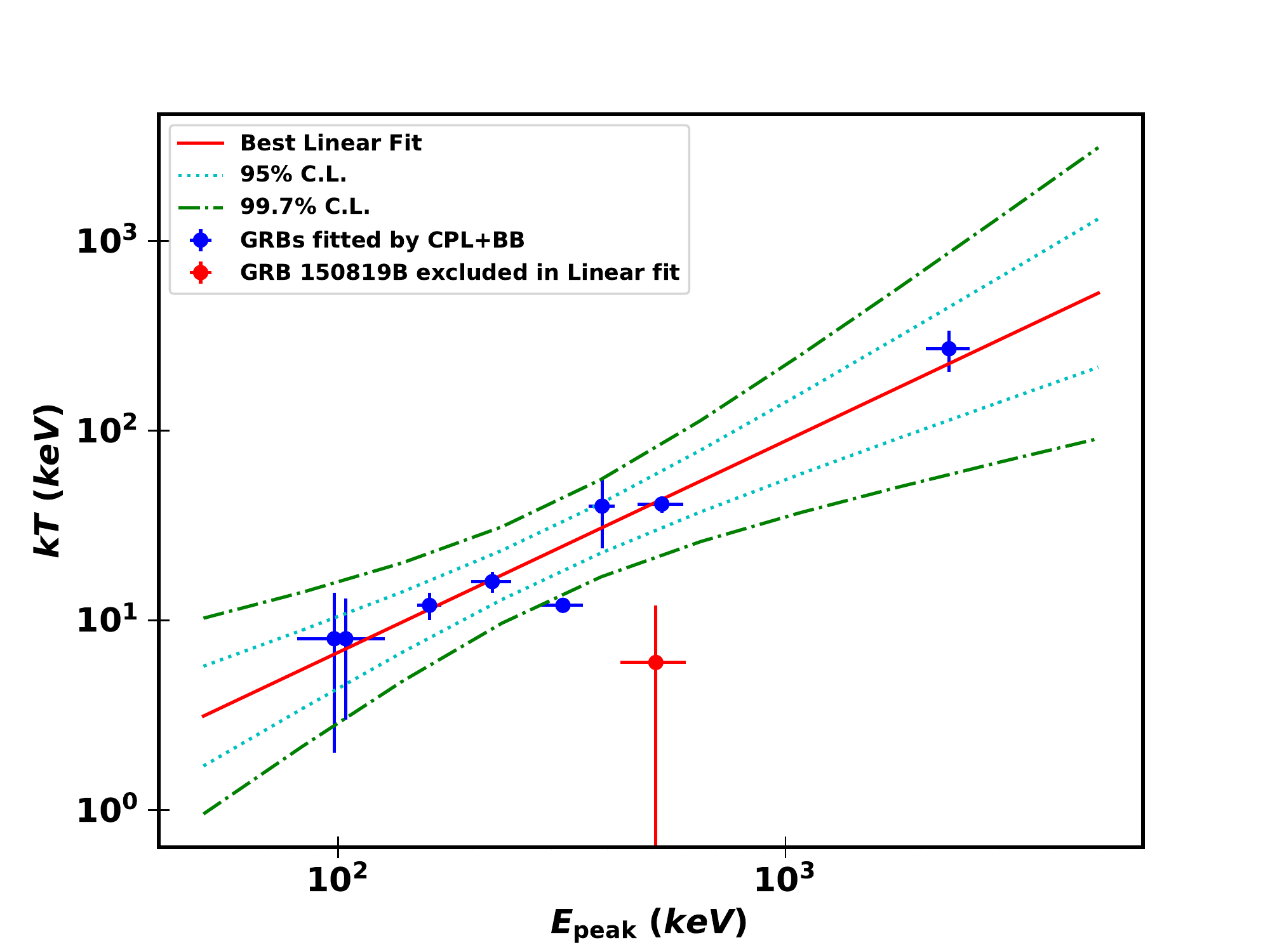}
	\caption{Correlation between peak energy $E_{peak}$ in CPL function and temperature $kT$ in the standard BB function when GRB spectrum is fitted by CPL + BB model.}
	\label{fig:spectrum}
\end{figure}

In order to compare the multiple spectral components, we select the CPL + BB model to test the correlation between the peak energy $E_{\rm peak}$ = $(2+\Gamma) E_c$ in the CPL component and the temperature $kT$ in the BB component using the linear fit in logarithmic space same as above.
Excluding GRB 150819B, whose median value of $kT$= 6 keV is less than 8 keV, we found that $E_{\rm peak}$ is strongly correlated with $kT$ with $R$ = 0.97, $m$ = $-1.40\pm0.31$ and \mbox{$n$ = $1.11\pm0.12$}.
The best fit for the correlation is written as
\begin{equation}
	log (E_{\rm peak}) = (-1.40\pm0.31) + (1.11\pm0.12) log(kT),
	\label{eq:correlation2}
\end{equation}
$p$ is about 1.0$\times$10$^{-4}$,  {which favors a strong correlation}, as seen in Figure ~\ref{fig:spectrum}. This strong correlation implies those two spectral components might share the same origins.

\section{Discussion \label{discussion}}

\subsection{Bias on the Transition Process of SGRBs}
As seen in previous sections, there are three types of SGRBs, such as with single-pulse SGRB, double-pulse SGRB and triple-pulse SGRB, which might be connected with different types of energy dissipation. Transition process within one impulsive explosion, that is from a fireball to a Poynting-flux-dominated outflow, has been found in several LGRBs, e.g., GRB 160625B and GRB 160626B~\citep{Zhang2018,Li2019,Li2020}. When the multiple-pulse SGRBs can be clearly separated, it could be the apparent evidence to constrain the transition times, such as the GRB 150819B and GRB 180703B in this work. Although such bias exists in individual SGRBs, the positive correlation between the FWHM and the rising time still hold in our sample, which might due to the very short $T_{90}$ duration and a small SGRB sample.  {This dispersion might be resolved when with more and more SGRBs detected in distinct types of pulses, which} could divide this correlation into two or more subclass.

\subsection{Implications for the Simultaneous Thermal and Non-Thermal Spectral Components}
More and more GRB spectra were  {discovered deviating} from the typical BAND model, while the model comprising a PL/CPL (nonthermal) and an additional BB (thermal) is found being fitted well in several GRBs~\citep{Abdo2009,Ackermann2010,Guiriec2010,Ackermann2011,Dainotti2018,Tang2021}. The nonthermal component was well described within the context of synchrotron radiation from particles in the jet, while the thermal component was interpreted by the emission from the jet photosphere, see ~\citep{Dainotti2018} and references therein. The PL component was claimed to originate most likely from the inverse-Compton process as well as the CPL component, whose seed photons are usually the synchrotron-induced photons. The high-energy exponential cutoff in the latter component (CPL) is naturally interpreted as annihilation between the high-energy photons and the low-energy photons~\citep{Tang2015,Chand2020}. The correlation found in this work, such as CPL's peak energy $E_{peak}$ is in proportion to BB's temperature $kT$ ($E_{peak} \propto kT^{1.1}$), would imply that  {despite} the different pulse-type SGRBs in our sample, the prompt broadband gamma-ray radiation could originate from the similar structures of the region,  {such as the same jet structures and outflow compositions in the different types of SGRBs. As seen in ~\citep{Tang2021}, both the leptonic model and hadronic model could produce the simultaneous thermal and non-thermal spectral components, more extensive physical explanations are required in~future.}

\section{Summary and Conclusions \label{conclusion}}
In this work, nine most bright SGRBs have been analyzed both in the LCs and SEDs. Of these bursts, the pulses in the LCs are all best-fitted by the FRED profiles. The resultant rising time width ($T_{\rm rise}$) is found to be strongly positive-correlated with the full time width at half maxima (FWHM), which implies those pulses are emitted within one impulsive explosion. This correlation might be divided into two or more subclass if more and more SGRBs in different pulse types are detected in future. The correlation of spectral parameter  found in this work, such as CPL's peak energy $E_{peak}$ is in proportion to BB's temperature $kT$ ($E_{peak} \propto kT^{1.1}$), would imply that  {despite} the different pulse-type SGRBs in our sample, the prompt broadband gamma-ray radiation could originate from the similar structures of the region.


\vspace{6pt}



\authorcontributions{ Writing---review and editing, P.-W.Z. and Q.-W.T. All authors have read and agreed to the published version of the manuscript.}

\funding{ {T.Q.W. was funded by National Nature Science Foundation of China grant numbers 11903017 and 12065017.}}

\institutionalreview{Not applicable.}

\informedconsent{Not applicable.}

\dataavailability{Data available on reasonable request to the authors.}


\conflictsofinterest{The authors declare no conflict of interest.}


\appendixtitles{no} 

\reftitle{References}




\begin{adjustwidth}{-\extralength}{0cm}

\end{adjustwidth}

\end{document}